\documentstyle[epsf]{elsart}
\begin{document}

\begin{frontmatter}
\rightline{HD-THEP-97-07}
\rightline{JLAB-THY-97-16}
\vskip 1truecm
\title{Non-perturbative dynamics of the heavy-light quark system 
       in the non-recoil limit}
\author{N. Brambilla$^{~1}$ $^*$ and A. Vairo}
\footnote{Alexander von Humboldt Fellow}
\address{Institut f\"ur Theoretische Physik, Universit\"at Heidelberg \\
         Philosophenweg 16, D-69120 Heidelberg, FRG\\and\\
          Jefferson Laboratory \\
         12000 Jefferson Ave., Newport News, VA 23606, USA \\ and\\
         Nuclear/High Energy Physics (NuHep) Research Center, \\
         Hampton University, Hampton, VA 23668, USA, \\
         $^*$  I.N.F.N., Sezione di Milano\\
         Via Celoria 16, 20133 Milano }

\begin{abstract}
Starting from the relativistic gauge-invariant quark-antiquark 
Green function we obtain the relevant interaction in the one-body limit, 
which can be interpreted as the kernel of a non-perturbative Dirac equation. 
We study this kernel in different kinematic regions, reproducing, 
in particular, for heavy quark the potential case and sum rules results. 
We discuss the relevance of the result for heavy-light mesons 
and the relation with the phenomenological Dirac equations used 
up to now in the literature. 
\end{abstract}

\end{frontmatter}

\newpage        

\pagenumbering{arabic}
  
\section{Introduction}

In the last years a lot of efforts has gone into the study of systems 
involving at least one heavy quark ($Q$). On the experimental side there are 
a lot of data on the heavy mesons states ($Q \bar{Q}$), 
new data already measured \cite{data} and a great expectation 
for the ones to come on the heavy-light states ($q\bar{Q}$).
On the theoretical side the situation is the following. The dynamics 
of the systems composed by two heavy quarks is quite well understood in terms
of potential interaction (static and relativistic corrections)
\cite{Wilson,BCP,DoSi,DQCD,BV,Soto,Oakes} obtained from the semirelativistic
reduction of the QCD dynamics (for lattice studies see \cite{NRQCD}).
In the heavy-light case it turns out useful to take advantage of the 
heavy quark symmetries \cite{isgurw}. Heavy-quark symmetry implies that, 
in the limit where $m_Q \gg  \Lambda_{\rm QCD}$, the long-distance physics 
of several observables is encoded in few hadronic parameters, which 
in general can be defined in terms of operator matrix elements in heavy quark 
effective theory (HQET). For some recent reviews we refer the reader to 
\cite{hqet}. In this framework a systematic expansion can be done 
in the small parameters $\Lambda_{\rm QCD}/ m_Q$ and $\alpha_{\rm s}(m_Q)$. 
In the limit $m_{Q} \to \infty$ all the physics can be expressed in terms 
of a small number of form factors depending  on the light quark and gluon 
dynamics only. Then, in HQET  the heavy-light  meson mass is given by
\begin{equation}
m_{M}= m_{Q}+ \bar{\Lambda} 
+ O \left( {1\over m_Q} \right) \, \,{\rm corrections}
\label{eq:hqun}
\end{equation}
The parameter $\bar{\Lambda}$ represents contributions coming
from all terms independent of the heavy-quark mass $m_Q$; it is one
of the non-perturbative parameters of the HQET, which have a similar 
status as the vacuum condensates in sum rules and QCD phenomenology. 
$\bar{\Lambda}$ can be fixed on the data, its actual calculation 
however needs a dynamical input. Of course the dynamics of the 
light quark is inerehently non-perturbative. Some approaches resort 
to dynamical calculation via phenomenological potential models 
\cite{hqetpot,dirpot}, sum rules \cite{neubert} or relativistic 
phenomenological equations \cite{amerdirac}. To have a well founded 
calculation of $\bar{\Lambda}$ is of great importance since its value 
affects the determination of many phenomenological quantities 
(cf e. g. \cite{bigi}).

In this letter we address the question of calculating the non-recoil 
corrections to the heavy-light mesons ($\bar{\Lambda}$) 
via a Dirac equation justified by the QCD dynamics. 
Our starting point is the quark-antiquark gauge-invariant 
Green function taken in the infinite mass limit of one particle. 
The only dynamical assumption is on the behaviour of the 
Wilson loop. The gauge invariance of the formalism guarantees  
that the relevant physical information are preserved at any step 
of our derivation. In this way we obtain a QCD justified 
fully relativistic interaction kernel for the quark 
in the infinite mass limit of the antiquark. 
This kernel reduces in some region of the physical parameters 
to the heavy quark mass potential, and leads in some other region 
to the heavy quark sum rules results, providing in this 
way an unified description. In the light of our result 
we scrutinize the phenomenological Dirac equations used 
in the literature and give an answer to the old-standing problem of the 
Lorentz structure of the Dirac kernel for a confining interaction
\cite{amerdirac,simdirac,olsdirac,sucher}.

There are many possible applications of the obtained result, 
like the study of relativistic properties of the spectrum 
(as much relevant as the quark is light) and the calculation 
of heavy-light meson matrix elements and form factors (e. g. the Isgur--Wise 
function). Finally, this work can also be intended as a step forward both  
in the direction of a theory derived two-body relativistic interaction, 
both in  the direction of a  generalization of the sum rules approach 
with the inclusion of a finite gluon correlation length. 

\section{The one-body interaction}

\begin{figure}[htb]
\vskip 0.8truecm
\makebox[2.5truecm]{\phantom b}
\epsfxsize=10truecm
\epsffile{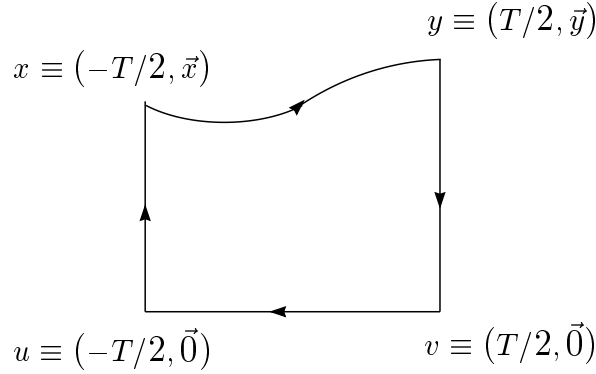}
\vskip -4truecm
\caption{{{\it The Wilson loop in the static limit of the heavy quark.}}}
\label{figwilsonh}
\vskip 0.8truecm
\end{figure}

The quark-antiquark Green function is given in quenched approximation by 
\begin{equation}
G_{\rm inv}(x,u,y,v) =  
\left\langle {\rm Tr} \, i\,S^{(1)}(x,y;A) U(y,v) \, 
i\,S^{(2)}(v,u;A) U(u,x)\right\rangle, 
\label{Ginv}
\end{equation}
where the points $x,y,u,v$ are defined as in Fig. \ref{figwilsonh},  
$\langle~~\rangle$ means the norma\-lized average over the gauge field  
$A_\mu$,  $S^{(i)}$ is the fermion propagator in the external field $A_\mu$ 
associated with the particle $i$ and the strings $U(y,x) \equiv {\rm P}\,
\exp \left\{ \displaystyle ig\int_0^1 ds\, (y-x)^\mu  A_\mu(x + s(y-x)) 
\right\}$ are needed in order to have gauge invariant initial 
and final bound states. A very convenient way to represent it 
is the so-called Feynman--Schwinger representation 
(see \cite{sitj,fsbs97} and refs. therein), where the fermion propagators 
are expressed in terms of quantomechanical path integrals over the quark 
trajectories ($z_1(t_1)$ and $z_2(t_2)$) 
\begin{eqnarray}
&~&G_{\rm inv}(x,u,y,v) =
{1\over 4} \Bigg\langle {\rm Tr}\,{\rm P}\, 
(i\,{D\!\!\!\!/}_{y}^{\,(1)}+m_1)\, 
\int_{0}^\infty dT_1\int_{x}^{y}{\cal D}z_1
e^{\displaystyle - i\,\int_{0}^{T_1}dt_1 {m^2+\dot z_1^2 \over 2}   }
\nonumber\\
&~&\qquad\quad\times
\int_{0}^\infty dT_2\int_{v}^{u}{\cal D}z_2
e^{\displaystyle - i\,\int_{0}^{T_2}dt_2 {m^2+\dot z_2^2 \over 2}   }
e^{\displaystyle ig \oint_\Gamma dz^\mu A_\mu(z)}
\nonumber\\
&~&\qquad\quad\times 
e^{\displaystyle i\,\int_{0}^{T_1}dt_1 {g\over 4}\sigma_{\mu\nu}^{(1)}
F^{\mu\nu}(z_1)}
e^{\displaystyle i\,\int_{0}^{T_2}dt_2 {g\over 4}\sigma_{\mu\nu}^{(2)}
F^{\mu\nu}(z_2)} 
(-i\,\buildrel{\leftarrow}\over{D\!\!\!\!/}_{v}^{\,(2)} + m_2) \Bigg\rangle . 
\label{Ginv2}
\end{eqnarray} 
From Eq. (\ref{Ginv2}) it emerges quite manifestly that the entire dynamics 
of the system depends on the Wilson loop:
\begin{equation}
W(\Gamma;A) \equiv   
{\rm Tr \,} {\rm P\,} \exp \left\{ ig \oint_\Gamma dz^\mu A_\mu (z) \right\},
\label{wilson}
\end{equation}
being  $\Gamma$ the closed curve defined by the quark trajectories 
and the endpoint strings  $U(y,v)$ and $U(u,x)$.

Let us assume that the antiquark moving on the second fermion line 
becomes infinitely heavy. The only trajectory surviving 
in the path integral of Eq. (\ref{Ginv2}) associated with the second 
particle is the static straight line propagating from $v$ to $u$. 
The corresponding Wilson loop of the system is represented in 
Fig. \ref{figwilsonh}. As already noted in \cite{bal85} in this case 
it turns out to be very convenient to choose the following gauge 
condition (sometimes called modified coordinate gauge):
$$
A_\mu(x_0,{\bf 0}) = 0, \qquad\qquad x^jA_j(x_0,{\bf x}) = 0 .
$$ 
Thanks to this gauge choice it is possible to express the gauge 
field in terms of the field strength tensor, 
\begin{eqnarray}
A_0(x) &=& \int_0^1 d\alpha\, x^k F_{k0}(x_0, \alpha {\bf x}), \nonumber\\
A_j(x) &=& \int_0^1 d\alpha\, 
\alpha \,x^k F_{kj}(x_0, \alpha {\bf x}) . \nonumber
\end{eqnarray}
Moreover the only non-vanishing contribution to the Wilson loop 
is given by the quark paths connecting $x$ with $y$, and we have 
\begin{equation}
W(\Gamma;A) =    
{\rm Tr \,} {\rm P\,} \exp \left\{ ig \int_x^y dz^\mu A_\mu (z) \right\}.
\label{wilson1}
\end{equation}
We stress that the choice of the gauge is in this approach really 
arbitrary and motivated only by convenience. Being the formalism 
completely gauge invariant, by handling properly 
we would obtain exactly the same results within any gauge.

\begin{figure}[htb]
\vskip 0.8truecm
\makebox[4.5truecm]{\phantom b}
\epsfxsize=6truecm
\epsffile{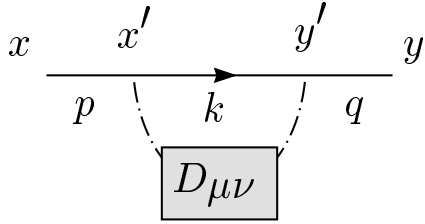}
\vskip 0.3truecm
\caption{{{\it The interaction kernel $K$.}}}
\label{figfockh}
\vskip 0.8truecm
\end{figure}

As showed in \cite{sitj,fsbs97} in order to evaluate Eq. (\ref{Ginv2}) 
we need to know the Wilson loop average over the gauge fields. 
We evaluate it via the cumulant expansion described in \cite{DoSi}. 
Keeping only bilocal cumulants we obtain:
\begin{eqnarray}
\langle W(\Gamma,A) \rangle &=& 
\exp \left\{ - {g^2\over 2} \int_x^y dx^{\prime\mu} \int_x^y dy^{\prime\nu}
D_{\mu\nu}(x^\prime,y^\prime) \right\},\nonumber\\
D_{\mu\nu}(x,y) &\equiv& 
x^ky^l\int_0^1 d\alpha \, \alpha^{n(\mu)} \int_0^1 d\beta\, \beta^{n(\nu)}
\langle F_{k\mu}(x^0,\alpha{\bf x})F_{l\nu}(y^0,\beta{\bf y})\rangle , 
\label{svm}
\end{eqnarray}
where $n(0) = 0$ and $n(i) = 1$. 
Assumption (\ref{svm}) corresponds to the so-called stochastic vacuum model 
and has been very successful in the last years either in applications to 
potential models as well as in the study of soft high energy scattering 
problems (for some recent reviews see \cite{do}). 
Inserting expression (\ref{svm}) in Eq. (\ref{Ginv2}) 
and expanding the exponential we obtain the following
expression for the propagator $S_D$ of the quark ($S_D$ is 
$G_{\rm inv}$  ``projected'' on the first fermion line; the second 
quark is irrelevant in the infinite mass limit, playing the role of 
an external source):
\begin{equation}
S_D = S_0 + S_0 \, K \, S_0 + S_0 \, K \, S_0 \, K \, S_0 + \cdots.  
\label{sdexp}
\end{equation} 
$S_0$ is the free fermion propagator. Taking into account only the first 
planar graph (since we are interested only in contributions proportional to 
the gluon condensate), we have $K(y^\prime,x^\prime) = \gamma^\nu 
S_0(y^\prime,x^\prime) \gamma^\mu D_{\mu\nu}(x^\prime,y^\prime)$. 
A graphical representation of $K$ is given in Fig. \ref{figfockh}.
Eq. (\ref{sdexp}) can be written in closed form  
as $S_D = S_0 + S_0 K S_D$ (or in terms of wave-function, 
$({p\!\!\!/} -m - iK)\psi = 0$; $m\equiv m_1$). Therefore, $K$ can 
be interpreted as the interaction kernel of the Dirac equation associated 
with the motion of a quark in the field generated by an 
infinitely heavy antiquark. 

Assuming that the correlator $\langle F_{\mu\lambda}(x)F_{\nu\rho}(y)\rangle$ 
depends only on the difference between the coordinates, we define: 
$$
\langle F_{k\mu}(x^0,\alpha{\bf x})F_{l\nu}(y^0,\beta{\bf y})\rangle \equiv
f_{k\mu l\nu}(x^0-y^0, \alpha{\bf x} - \beta{\bf y}). 
$$
With this assumption $K$ can be written in momentum space 
as (see Fig. \ref{figfockh} for the definition of the momenta): 
\begin{eqnarray}
K(q,p) &=& -g^2(2\pi)\delta(p^0-q^0) \nonumber\\
&\times& 
\int_{-\infty}^{+\infty}d\tau \int_0^1 d\alpha \, \alpha^{n(\mu)}
\int_0^1 d\beta \, \beta^{n(\nu)} 
{\partial\over\partial p^k}{\partial\over\partial q^l}
\int d^3r \,e^{i({\bf p}-{\bf q})\cdot{\bf r}} \nonumber\\
&\times& 
\gamma^\nu\left\{ \theta(-\tau)\Lambda_+({\bf t})\gamma^0 e^{-i(p^0-E_t)\tau}
- \theta(\tau)\Lambda_-({\bf t})\gamma^0 e^{-i(p^0+E_t)\tau} \right\}\gamma^\mu
\nonumber\\
&\times&
f_{k\mu l\nu}(\tau, (\alpha - \beta){\bf r}),
\label{kmom}
\end{eqnarray}
where ${\bf t} \equiv (\beta{\bf p}-\alpha{\bf q})/(\beta-\alpha)$, 
$E_t = \sqrt{t^2+m^2}$ and $\Lambda_{\pm}({\bf t}) = 
{\displaystyle{E_t \pm (m-{\bf t}\cdot\gamma)\gamma^0 \over 2 E_t}}$.

Equation (\ref{kmom}) is our basic expression. It contains the 
perturbative interaction up to order $g^2$ and the non-perturbative 
one carried by a single insertion of a second order cumulant. 
From now on we want to focus our attention only on the 
purely non-perturbative interaction. The Lorentz structure of the 
non-perturbative relevant part of $f_{\mu\lambda\nu\rho}$ is  
\begin{equation}
f^{\rm n.p.}_{\mu\lambda\nu\rho}(x) = {1 \hskip -2.2 pt {\rm l}}_{\rm c} \,
{\langle F^2(0) \rangle \over 24 N_c}
(g_{\mu\nu} g_{\lambda\rho} - g_{\mu\rho} g_{\lambda\nu}) D(x^2),
\label{ddd}
\end{equation}
where  $\langle F^2(0) \rangle$ is the gluon condensate, 
${1 \hskip -2.2 pt {\rm l}}_{\rm c}$ the identity matrix of SU(3) and 
$D$ is a non-perturbative form factor normalized to unit at the origin.
Lattice simulations have showed that $D$ falls off exponentially 
(in Euclidean space-time) at long distances with a correlation 
length $a^{-1} \sim (1 ~{\rm GeV})^{-1}$ 
\cite{lat}. This behaviour of $D$ is sufficient 
to give confinement, at least in some kinematic regions \cite{DoSi}. 
Moreover we notice that, if $f^{\rm n.p.}_{\mu\lambda\nu\rho} \sim 
{\displaystyle e^{-ia\tau}}$, then  we have 
$$
K \sim {\partial\over\partial p^k}{\partial\over\partial q^l}
\int d^3r \, e^{i({\bf p}-{\bf q})\cdot{\bf r}} 
\gamma^\nu S_0(p_0+a,{\bf t}) \gamma^\mu 
f^{\rm n.p.}_{k\mu l\nu}(0, (\alpha - \beta){\bf r}).  
$$
The main effect of the finite correlation length $a^{-1}$ seems to 
consist, therefore, in a shifting of the pole in the inserted 
free fermion propagator. 

In the following we will study expression (\ref{kmom})   
for different choices of the parameters which are 
the correlation length $a$, the mass $m$, the binding energy 
$(p_0 - m)$ and the momentum transfer $({\bf p} - {\bf q})$. 
 
{\bf A. Heavy quark potential case}  ($m>a>p_0-m$)

If we assume $a$ to be bigger than the binding energy $(p_0 - m)$ 
and smaller than the mass $m$ of the quark, 
since $a\sim 1$ GeV, the quark turns out to be sufficiently heavy 
to be considered non-relativistic. In order to obtain the $1/m^2$ 
potential we can neglect the ``negative  energy states'' 
contributions to (\ref{kmom}) by writing  
\begin{eqnarray}
K(q,p) &\simeq& -g^2(2\pi)\delta(p^0-q^0) \nonumber\\
&\times& 
\int_0^{+\infty}d\tau \int_0^1 d\alpha \, \alpha^{n(\mu)}
\int_0^1 d\beta \, \beta^{n(\nu)} 
{\partial\over\partial p^k}{\partial\over\partial q^l}
\int d^3r e^{i({\bf p}-{\bf q})\cdot{\bf r}} \nonumber\\
&\times& 
\gamma^\nu \Lambda_+({\bf t})\gamma^0 \gamma^\mu 
f^{\rm n.p.}_{k\mu l\nu}(\tau, (\alpha - \beta){\bf r}). 
\label{kmomA}
\end{eqnarray}
Now, inserting Eq. (\ref{ddd}) and by means of usual reduction 
techniques \cite{report}, we obtain up to order $1/m^2$ the static and 
spin dependent potential 
\begin{eqnarray}
V(r) &=& g^2 {\langle F^2(0) \rangle \over 24 N_c} 
\int_{-\infty}^{+\infty} d\tau \int_0^r d \lambda \, (r-\lambda)\,  
D(\tau^2 - \lambda^2) \nonumber\\
&+& {{\bf \sigma}\cdot {\bf L} \over 4 m^2} {1\over r}
g^2 {\langle F^2(0) \rangle \over 24 N_c} 
\int_{-\infty}^{+\infty} d\tau \int_0^r d \lambda 
\left( 2{\lambda\over r} - 1\right) D(\tau^2 - \lambda^2),
\label{pot}
\end{eqnarray}
where we have make use of the change of variable 
$\lambda = (\alpha - \beta)|{\bf r}|$. 
This result agrees with the one body limit of the potential 
given in \cite{DoSi,BV}. In particular for $r\to\infty$ 
identifying the string tension $\sigma = \displaystyle g^2 
{\langle F^2(0) \rangle \over 24 N_c} \int_{-\infty}^{+\infty} d\tau  
\int_0^\infty d \lambda \, D(\tau^2 - \lambda^2)$ 
we obtain the well-known Eichten and Feinberg result \cite{Wilson}, 
\begin{equation} 
V(r) = \sigma r -C
- {{\bf \sigma}\cdot {\bf L} \over 4 m^2} \,\, {\sigma \over r}, 
\label{EF}
\end{equation}
where $C$ is the constant term $\displaystyle g^2 {\langle F^2(0) 
\rangle \over 24 N_c} \int_{-\infty}^{+\infty} d\tau  \int_0^\infty d \lambda 
\, \lambda\, D(\tau^2 - \lambda^2)$. 
We observe that the Lo\-rentz structure  which gives origin to the negative 
sign in front of the spin-orbit potential in (\ref{EF}) is in our 
case not simply a scalar ($K\simeq \sigma \, r$). We will discuss this 
point in more detail in the conclusions. 

{\bf B. Sum rules case}  ($a<p_0-m$, $a<m$)

Let us consider now the case in which the binding energy of the quark is 
bigger than the correlation length, which can be considered zero respect 
to all the scales of the problem. In the literature is usually referred 
to this case as the non potential case \cite{DoSi}.  
Since $f^{\rm n.p.}_{\mu\lambda\nu\rho}(x) {\displaystyle 
{\mathop{\longrightarrow}\limits_{a\to 0}} } 
f^{\rm n.p.}_{\mu\lambda\nu\rho}(0)$ $= {1 \hskip -2.2 pt {\rm l}}_{\rm c} 
\langle F^2(0) \rangle /24 N_c
(g_{\mu\nu} g_{\lambda\rho} - g_{\mu\rho} g_{\lambda\nu})$, we have   
\begin{eqnarray}
K(q,p) &\simeq& -g^2 (2\pi)\delta(p^0-q^0)
{1 \hskip -2.2 pt {\rm l}}_{\rm c}{ \langle F^2(0) \rangle \over 24 N_c} 
(g_{\mu\nu} g_{kl} - g_{\mu l} g_{\nu k}) 
\nonumber\\
&\times& 
\int_0^1 d\alpha \, \alpha^{n(\mu)}
\int_0^1 d\beta \, \beta^{n(\nu)} 
{\partial\over\partial p^k}{\partial\over\partial q^l}
\left( \gamma^\nu S_0(p) \gamma^\mu (2\pi)^3\delta^3({\bf p}-{\bf q})\right). 
\label{kmomB}
\end{eqnarray}
In particular from Eq. (\ref{kmomB}) we obtain the well-known leading 
contribution to the heavy quark condensate \cite{svz}:
$$
\langle \bar{Q} Q \rangle  = 
-\int {d^4p \over (2\pi)^4}\int {d^4q \over (2\pi)^4} 
{\rm Tr} \left\{ S_0(q)K(q,p)S_0(p) \right\} = -{1\over 12} 
{\langle \alpha F^2(0) \rangle \over \pi m}.
$$ 

{\bf C. Light quark case}  ($a>m$)

Since we have reproduced the known results concerning heavy quarks, 
Eq. (\ref{kmom}) should maintain some physical meaning also by considering 
heavy-light mesons with a strange quark (like D$_{\rm s}$ and  B$_{\rm s}$). 
In this case the light quark mass is smaller than $a$: $m_{\rm s} \sim$ 
200 MeV $< 1$ GeV. Actually the case $a>m$ 
has to be considered as the only realistic one concerning 
heavy-light mesons. Under this condition either the 
exponent $(p_0 - E_t)$ as well as $(p_0 + E_t)$ can be 
neglected with respect to $a$. Therefore we have:
\begin{eqnarray}
K(q,p) &\simeq& -g^2 (2\pi)\delta(p^0-q^0) \nonumber\\
&\times& 
\int_0^{+\infty}d\tau \int_0^1 d\alpha \, \alpha^{n(\mu)}
\int_0^1 d\beta \, \beta^{n(\nu)} 
{\partial\over\partial p^k}{\partial\over\partial q^l}
\int d^3r e^{i({\bf p}-{\bf q})\cdot{\bf r}} \nonumber\\
&\times& 
\gamma^\nu\left( {m-{\bf t}\cdot\gamma \over E_t} \right) \gamma^\mu
f^{\rm n.p.}_{k\mu l\nu}(\tau, (\alpha - \beta){\bf r}). 
\label{kmomC}
\end{eqnarray}
We observe, as an appealing feature of this expression, 
that in the zero mass limit it gives a chirally symmetric 
interaction (while a purely scalar interaction breaks chiral symmetry 
at any mass scale). On the other hand, by considering only its static 
contribution (i. e. neglecting all momentum dependent terms), 
we obtain (using the same definition of the string tension given previously): 
$$
K(q,p)\big|_{\rm static} =  g^2 (2\pi)\delta(p^0-q^0)\,
\int d^3r e^{i({\bf p}-{\bf q})\cdot{\bf r}} 
{5\over 3} \, \sigma \, r. 
$$ 
The factor $5/3$ in front of $\sigma$ which arises naturally in this approach 
under the considered physical conditions, seems to supply 
an explanation for the fact, observed by many authors 
\cite{amerdirac,simdirac,olsdirac}, that in order to reproduce the D$_{\rm s}$ 
and B$_{\rm s}$ spectra from a Dirac equation with scalar confinement a string 
tension almost twice respect to the usual value is needed. 
 
\section{Conclusions}

In the literature, also recently, a Dirac equation 
with scalar confining kernel (i. e. $K \simeq \sigma \,r$) 
has been used in order to calculate  non-recoil contributions 
to the heavy-light meson spectrum \cite{amerdirac,simdirac,olsdirac}. 
The main argument in favor of this type of kernel is the nature of the 
spin-orbit potential for heavy quarks. This turns out to have a long-range 
vanishing magnetic contribution (according to the Buchm\"uller 
picture of confinement) and is completely described by the Thomas 
precession term. This situation is compatible with a scalar confining  
kernel. However, assuming more sophisticated confinement models with a bigger 
sensitivity to the intermediate distance region, the spin-orbit 
interaction has no more such a simple behaviour. In particular 
there show up  non zero corrections to the magnetic spin orbit 
potential. Moreover, the velocity dependent sector 
of the potential seems not to be compatible with a scalar kernel 
(we refer the reader to \cite{BV} for an exhaustive discussion). 
Therefore also from the point of view of the potential 
theory there are strong indications that the Lorentz structure of the 
confining kernel should be more complicate that a simple scalar one. 
This emerges also in our approach. The kernel (\ref{kmom}) follows 
simply from the assumption on the gauge fields dynamics given 
by Eq. (\ref{svm}). In principle all the graphs constructed by 
inserting non-perturbative gluon propagators on the quark fermion line 
should be taken into account. Since we are interested only 
in terms proportional to $\sigma$ (or $\langle F^2(0)\rangle$), 
we keep only the first one. When performing the potential reduction of this 
kernel in the heavy quark case (A) we obtain exactly 
the expected  static and spin-dependent  potentials. 
Therefore our conclusion is that there exists at least one 
non scalar  kernel which reproduces for heavy quark 
not only the Eichten and Feinberg potentials in the long distances limit, 
but also the entire stochastic vacuum model spin dependent potential.  
Moreover when considering $a$, the inverse of the correlation length, small 
with respect to all the energy scales (case B), the kernel (\ref{kmom}) 
gives back the leading heavy quark sum rules results.  
It is possible to try to extend the range of applicability of 
Eq. (\ref{kmom}) to more realistic cases, like D$_{\rm s}$ and 
B$_{\rm s}$ mesons where the light quark mass is smaller than the 
cha\-racteristic correlation length of the two point cumulant (case C). 
The relevant part of the kernel is also in this case not a simply scalar 
one. In the static approximation this kernel seems, indeed, 
to be compatible with the existing phenomenology.  
We notice, however, that the situation is quite different from QED where 
in the Coulomb gauge a static interaction emerges without any approximation  
since the transverse part of the propagator of the exchanged photon 
(the relevant contribution to the binding) vanishes 
when the second fermion line is taken infinitely heavy. 
In our approach a purely static contribution never emerges, being 
relevant to the binding not an exchange graph  between 
the quarks, but the interaction with the background vacuum fields. 
Therefore any static approximation in a real heavy-light system,  
for which the light quark is expected to be far from a static one,  
appears doubtful. 

\vskip 1truecm
{\bf Acknowledgments} 
The authors gratefully acknowledge the Jefferson Lab Theory Group as well as 
the Hampton University for their warm hospitality during the first 
stage of this work, the Alexander von Humboldt 
Foundation for support and D. Gromes for useful discussions. 
This work was supported in part by NATO grant under contract N. CRG960574.

\end{document}